Corresponding Author: Prof. muller pierre, phD

Corresponding Author's Institution:

First Author: muller pierre, phD

Order of Authors: muller pierre, phD; Ranguelov Bogdan, PhD; Metois Jean Jacques, PhD

# Spirals on Si(111) at Sublimation and Growth REM and LODREM Observations

B. Ranguelov[1], J.J. Métois[2], P. Müller[2,3]

[1] Institute of Physical Chemistry – Bulgarian Academy of Sciences, "Acad. G. Bonchev"Str., bl. 11, 1113- Sofia, Bulgaria

[2] CRMCN/CNRS associated to Universities Aix Marseille II and Aix Marseille III, Campus de Luminy F-13288, Marseille, France

3 also Université Paul Cézanne

**Abstract**

Using recently proposed improvements of Reflection Electron Microsopy (REM) we study in perfectly controlled thermodynamics conditions spiral shapes and spirals on Si(111) surface. It is shown that the new method named low distortion reflection electron microscopy (LODREM) is a powerful instrument, resolving in much more details (compared with REM) growth or evaporation spirals at the crystal surface. More precisely, we examine the distance between two successive steps of a spiral at growth (or evaporation) with respect to the supersaturation (or undersaturation). It is found that this distance scales with an exponent close to -1/2. This result, which deviates from the BCF theory originates from a non local behavior with a slow kinetic of attachment of the adatoms at the steps.

**Introduction**

Recently, two main improvements of Reflection Electron Microscopes (REM) have been proposed [1,2]. The first one concerns the control of the thermodynamic conditions [1]. Indeed, thank to the glancing conditions used in REM, it has been possible to install in the sample holder a second sample. The two samples can be heated independently so that by tuning the current through the two samples, growth, evaporation or adsorption/desorption steady state or equilibrium conditions can be perfectly controlled. The second main improvement concerns the image processing [2]. Indeed, because of the glancing conditions, REM images are severely foreshortened in one direction so the use of REM for studying the shape of 2D objects remained a challenge. Recently, we proposed a simple modification of the microscope column (based on controlled tilts of the screen) to obtain non distorted images [2]. The so-obtained images are now available for shape analysis so that we did not hesitate to give a new acronym to the technique: LODREM for Low Distortion Reflection Electron Microscope [2]. In this paper, we will use a combination of both improvements for studying growth then evaporation spirals in perfectly defined thermodynamic conditions. More precisely we will give some results concerning the spiral shapes and the spiral behavior during growth and sublimation. In particular, we will show that the distance between spiral steps deviates from the standard BCF model [3] but can be interpreted on the basis of more recent predictions [4] that take into account the non instantaneous kinetic of attachment of the adatoms to the steps.

## I. Experimental setup

The experimental setup has been partially described in previous papers [10,11], nevertheless we will describe with more details than in [1] the sample carrier used to tune the thermodynamic conditions

- The silicon wafer we use are of type p (B) (resistivity $\rho \approx 1 \Omega\ cm$. The samples (20.0 x 2.0 x 0.3 mm$^3$) after standard cleaning procedure in acetone and ultrasonic bath are introduced in the column of the UHV electron microscope with base pressure of 10$^{-9}$ Torr. Before performing the experiments we heat *in situ* the two samples to 1570 K for several minutes.

- The specimen holder carries two samples, which we will call substrate and source. The source, parallel to the substrate, is placed close (~100μm) to the substrate (see Fig.1). The samples can be heated independently so that by adjusting the currents through the two silicon samples it is possible to tune equilibrium, evaporation or growth conditions. More precisely when the temperature of the source $T_{source}$ is lower than the temperature of the substrate undersaturation conditions are fulfilled and the substrate is evaporating (sublimation) at a given temperature $T_{substrate}$. Increasing the undersaturation, (from negative values to zero) the equilibrium conditions are theoretically achieved when the source and the substrate have the same temperature. In fact, because of the finite size of the samples, some atoms leaving the source are lost and do not reach the substrate. The true experimental equilibrium condition thus does not correspond to equal temperatures. Since observing the substrate by electron microscopy, the true equilibrium conditions can be checked. It corresponds to conditions for which no movement of monoatomic steps on the surface is observed[1]. In our experimental conditions, the steady-state is reached (no steps movement) when the temperature of the source is around 30 degrees more than the substrate temperature. Increasing then the temperature of the source $T_{source} > T_{substrate} + 30K$ leads to supersaturation conditions, and obviously to the observation of the growth process (via step advancement) on the substrate surface.

Since the sample and the source are both silicon, it is possible to use the activation energy of silicon desorption E = 4.3 eV [5] to write the resulting deposited flux on the substrate under the form:

$$F_{res} = K \left( e^{-E/kT_{source}} - e^{-E/kT_{substrate}} \right) \qquad (1)$$

where $K$ is a constant. Obviously, neglecting the finite size of both samples, (1) gives the equilibrium conditions for $\tau = T_{source}/T_{substrate} = 1$ (saturation), growth conditions for $\tau > 1$ (supersaturation) and evaporation conditions for $\tau < 1$ (undersaturation). In Figure 2 is plotted $F_{res}$ versus the relative temperature $1/\tau = T_{substrate}/T_{source}$. In growth conditions and for a given source temperature $T_{source}$, $F_{res} > 0$ is quasi-independent upon $\tau$ that means that desorption from the substrate can be neglected against the incoming flux. In evaporation conditions $F_{res} < 0$ varies a lot with $\tau$. In a similar way, for a given $T_{substrate}/T_{source}$ ratio, in growth conditions $F_{res}$ weakly depends upon the source temperature while in sublimation conditions, it strongly depends upon it.

\*. In the Reflection Electron Microscope (REM) the electron beam, which passes between the two samples, is reflected at grazing incidence from the substrate. Because of the grazing incidence angle the image is foreshortened by roughly a factor 50 in the direction perpendicular to the electron beam [6]. This leads to an advantageous and an disadvantageous.

---

[1] Obviously evaporation (resp. growth) conditions correspond to atomic steps retreat (resp. advance).

The main advantageous is that the magnification distortion allows to observe large scales in one direction. It is very suitable for observing step motions on vicinal surfaces and especially for studying the step bunching phenomena at different temperatures. The main drawback is that the distortion does not allow to accurately resolve two-dimensional surface structures like growth or sublimation spirals emerging from screw dislocations [2,7]. This drawback can be avoided by using LODREM technique [2] whose basic idea lays in the controlled tilt of the screen, and thus overcomes the image distortion produced by the grazing incidence angle of the electrons on the surface. In spite of the partial lost of contrast due to the grazing angle of the incident electrons on the screen, LODREM appears to be an excellent *in-situ* technique for studying the dynamics between spirals, individual steps and step bunches, all three features well presented on Si(111) surfaces during growth or sublimation.

## II. Spirals at the surface a brief state of the art

Perfectly flat crystal faces are known to grow at relatively high supersaturations via two-dimensional nucleation. However it was soon observed, that, because of the presence of surface defects, crystal growth occured even at very low supersaturations. Frank [8] stressed the important role of the screw dislocations. Indeed since the seminal work of Burton, Cabrera and Frank [3], it is known that crystals grow by adatom incorporation at monoatomic steps. Thus a dislocation with a component of the Burgers vector normal to the surface is a step-source required for crystal growth. Then during growth, the step (pinned by the surface-emerging dislocation center) winds around the dislocation center and thus forms a spiral. The distance between two neighboring steps in a spiral thus appeared to be a pertinent parameter to describe the growth properties of a spiral. Let us consider in more details the inter-step distance in case of isotropic growth and at high temperatures where the spiral is rounded (not polygonized).

- In the classical BCF model of crystal growth [3], adatoms diffuse on the surface until desorption or incorporation in the closer step. The step velocity thus is completely determined by the local supersaturation $\sigma$ and the distance $\Lambda$ between two successive turns of a single arm spiral is shown to be $\Lambda = 4\pi r_c$, where $r_c$ is the radius of the critical nucleus. Cabrera and Levine [9] gave more accurate estimation of the distance between the spiral steps $\Lambda = 19 r_c = \frac{19 \kappa a^2}{kT\sigma}$, where $\kappa$ is the specific edge energy of the step and $a$ is the lattice parameter. Thus $\Lambda$ appears to be proportional to $\sigma^{-1}$.

- This classical BCF formulation is only valid for local step dynamic. Indeed, when taking into account the diffusion lengths of the adatoms $\lambda_s = \sqrt{D\tau_s}$, where $D$ is the diffusion coefficient and $\tau_s$ is the mean "life" time of an adatom on the surface, two different regimes appear. For $\lambda_s \ll \Lambda$ an adatom stays on the same terrace until incorporation in a step, the step dynamics thus is local, and the spiral growth regime well described by the previous paragraph. For $\lambda_s \gg \Lambda$ the diffusion fields from adjacent steps overlap so that successive turns of the spiral are strongly coupled via diffusion. The so-called "back force" or "back stress" effect was first considered by Cabrera and Coleman [10]: because of the dissymmetric diffusion field, the supersaturation at the spiral center is smaller, so that the radius of the first turn of the spiral is larger than the others. Obviously, the back force effect is not limited to the first turn [11]. One consequence of the back force effect is that at the limit $\lambda_s/\Lambda \to \infty$, far from the spiral center $\Lambda$ is constant and proportional to $\sigma^{-1/3}$.

- More recently, Karma and Plapp [4], revisit the spiral growth by means of a phase field model in which the assumption of instantaneous attachment of the adatoms to the steps has been released. In this case, they predict that for slow attachment kinetics the scaling is $\sigma^{-1/2}$, the Cabrera and Coleman scaling $\sigma^{-1/3}$ being only valid for instantaneous kinetics.

As a conclusion of this introductory section, one can retain that there is $\Lambda \propto \sigma^{-n}$ with $1/3 < n < 1$ for $\lambda_s \ll \Lambda$ but $1/2 < n < 1/3$ for $\lambda_s \gg \Lambda$.

### III. Experimental results and discussion

Our main experimental results concerning simple spiral observations are: (i) the spiral density we found roughly is in the range of $10^5$ to $10^4$ cm$^{-2}$. (ii) All of the spirals observed at the surface of Si(111) are single arm spirals, producing steps with monoatomic height. This is also confirmed by observing the step-step annihilation, produced by two neighboring spirals. (iii) We do not observe any cooperative spirals with more than one arm. (iv) The rotating clock- or counterclockwise spirals are found to be roughly numerically equal, irrespective to growth or sublimation conditions.

In this section we will give and comment the first LODREM images of spirals then we will be interested in the spiral step distance versus the supersaturation. At last, we will describe some dynamic behavior.

### *III.1. First LODREM images of spirals*

Let us consider some isolated spiral. The zone close to the center of the spiral is shown in Fig. 3., where images for four different angles of the screen are presented.

The first image on Fig. 3. is close to the conditions for REM observation , while the last one is almost at the maximum of the achievement of the LODREM technique with the screen tilt axis in the horizontal direction (in fact, because of a geometrical constraint due to the screen holder, there remains a residual distortion of 10 %). Again, let us underline that the LODREM image enables us to study in detail the morphology of the spiral whereas it is not possible in REM conditions (compare Fig. 3a to Fig. 3d) . In order to completely resolve the shape of the spiral at its center additional digital processing of this last image was performed until full elimination of the distortion. The result is shown in Fig. 4., where the digitally processed image is presented (left panel), and where we give a comparison between the shape of the spiral and an Archimedean spiral (right panel). It is seen that the spiral shape is very close to an Archimedean one.

Let us now consider the case of spirals interacting with steps, step bunches or other spirals.

On the first image (Fig. 5) two spirals (whose centers are materialized by black and white arrows) with reverse directions of rotation are close to each other. They are also close to a step bunch (black band in Fig. 5) which separates the spirals from a train of monoatomic steps (upper part of the image). Because of the spiral-spiral interaction, as well as because of the local slope due to the step bunch, the morphology of the spirals is strongly altered and thus deviates from the Archimedean shape of one individual spiral. A similar situation is presented on the second image of Fig. 5. where the spiral shape is affected by a train of individual steps close to a bunch.

### *III.2/ Dynamic behavior of the spiral steps :*

As written in the previous section, the initial state of our Si surface is a consequence of the high temperature annealing process, so that transitory states cannot be seen. In other words, we observe a surface with already formed spirals of sublimation, which are in fact "negative" spiral holes in the bulk emerging at the crystal surface. During sublimation the negative spirals holes become deeper and deeper, while during growth they fill. This gives birth to a very peculiar step behavior we will now describe.

(i) During sublimation the direction of movement of the steps is outward from the center of the spiral (see Fig.6). The monoatomics steps recede from the center to the periphery of the spiral and the hole becomes deeper and deeper.

(ii) Getting closer to equilibrium conditions (the saturation increasing from negative values towards zero), so that the inter-step distance increases (see next section and figure 8), the steps slowed down and at equilibrium stop to move.

(i) Switching to growth conditions (the saturation increases from zero to positive values) the direction of the movement of the steps switches from the outer region to the center (Fig. 9.). The monoatomic steps advance from the periphery to the spiral center and the hole is filling. Notice that we were able to observe this "switching" only by carefully maintaining the growth conditions and keeping them very close to equilibrium.

(ii) Of course, after certain time of deposition the hole is completely filled so that a flat surface should be recovered. However there is always a surface defect originating from the bulk dislocation and emerging at the surface. Thus the growth gives birth to a "positive" pyramidal spiral with obviously a new switch for the step movement since now the steps must advance form the center towards the periphery of the spiral.

Thus the direction of movement of the steps depend at once upon the saturation and upon the local geometry. Though quite obvious, it is the first time we directly observe this phenomenon.

*III.2. Distance between spiral steps on Si(111) in growth then sublimation conditions:*

In order to perform experiments in growth conditions on Si(111), we keep the substrate at a constant temperature and increase the temperature of the source. In other words, we increase the supersaturation $\tau = T_{source}/T_{substrate}$. The resultant flux $F_{res}$ (bilayers $d_{(111)}$=3.14 Å per second) is measured from the step velocity far away from the center of the spiral, where the curvature of the steps is negligible. Though, from a theoretical viewpoint, some time evolution of the step spacing during growth or sublimation should be expected [4,11], we do not observe such change. It is because the steady-state is reached before the end of the annealing process necessary for cleaning the sample. Thus we do not have access to the eventual transitory states and only give results concerning the steady state.

The results for four different substrate temperatures are presented in Fig.6.

It is seen (Fig. 7) that for all temperatures, the step distance scales with the deposition rate (variation of the deposition rate equal to two orders of magnitude) $\Lambda \propto \sigma^{-n}$ with a scale exponent n=$0.45 \pm 0.05$ [2].

We perform similar experiments for the case of evaporation on Si(111) at constant temperature and different values of the undersaturation. For this purpose we keep the substrate at a constant temperature and starting from near-equilibrium conditions, we decrease the flux of impinging atoms from the source, and thus increase the undersaturation until reaching the case of pure sublimation at the given temperature. The results for three different substrate temperatures are presented in Fig. 8.

Again like in the case of growth, the step distance scales with the sublimation rate $\Lambda \propto \sigma^{-n}$ with a scale exponent n= $0.40 \pm 0.05$.

The main difference between the figures 7 and 8 is that in the case of sublimation, the curves shift with the substrate temperature while in growth conditions they superimpose. This behavior can be easily understood when considering Fig. 2. For a given $T_{source}$, in growth conditions $F_{res}$ essentially does not depend on the substrate temperature since the desorption from the substrate can be neglected against the incoming flux. In evaporation conditions $F_{dep}$ essentially depends upon the substrate temperature since the leading term is the desorption from the substrate (which obviously depends upon the substrate temperature !). In other words, for growth conditions, the deposition flux essentially depends on the temperature of the source (we reach the value of approximately 6 – 7 bl/s the utmost limit we are able to measure) while for sublimation conditions the deposition flux essentially depends upon the substrate temperature. For the highest values of the undersaturation, the deposition flux is merely zero, and the regime of pure sublimation is reached.

---

[2] The value we give is an average from the four slopes measured at 1170°C, 1200°C, 1250 °C, 1270°C.

Thus we can conclude that in sublimation as well as in growth conditions we find $\Lambda \sim F^{-\alpha}$ between the distance $\Lambda$ and the deposition flux $F$ with $1/3<\alpha<1/2$ but closer to ½, at least for growth conditions.

Let us now discuss in more details our experimental conditions.

(iii) Our experiments take place at a temperature which is above the (7x7) → (1x1) transition. At such temperature, (i) the equilibrium concentration of adatoms on Si(111) is known to be extremely high (between one third and one quarter of a monolayer [12]) and (ii) the diffusion length of the adatoms $\lambda_s$ roughly varies from $10^6 \text{\AA}$ for 1100 °C to $10^5 \text{\AA}$ for 1250°C while according to the super or undersaturation conditions there is $10^4 \text{\AA} < \Lambda < 10^6 \text{\AA}$ (see Fig. 6 and 7) [13,14]. Thus, in our experimental conditions the "back-force" effect plays so that the usual BCF picture $\Lambda \propto \sigma^{-n}$ with n=1 is no longer valid. Thus, referring to section III there must be 1/3<n<1/2.

(iv) The experiments have been performed in a temperature range where the steps are reputed to be transparent [15-17]. In other words, because of the low density of kinks, most of the adatoms which diffuse on the terraces do not incorporate the steps but cross over the steps and thus visit several terraces prior to incorporation. The direct hopping of adatoms across the steps thus causes a new kind of coupling between the diffusion fields of neighboring terraces (see for example [18-20]) that slows down the kinetics. In such a case of slow kinetics, Karma and Plapp [4] have predicted $\Lambda \propto \sigma^{-n}$ with n=1/2 close to what we have found. Let us underline, that to the best of our knowledge, it is the first time that this prediction is experimentally verified. However, to be complete, let us note that scaling behavior $\Lambda \propto \sigma^{-n}$ with n=1/3 have been reported [21] for the case of MBE growth of PbTe on BaF$_2$(111) substrates.

Let us now focus on the weak scaling difference between growth (n=$0.45 \pm 0.05$) and evaporation (n=$0.40 \pm 0.05$) whose meaning should be that the kinetics of attachment is slower in growth conditions than in sublimation conditions. However if, at low temperature (T<900K), non equilibrium kinks are known to be responsible for the decrease of the step transparency (and thus the increase of the attachments kinetics) versus the incoming flux, it is not the case at higher temperature where the transparency properties become flux independent [17], so that whatever the temperature the step transparency cannot be more important in growth than in sublimation conditions. Another origin of this discrepancy could be found in the recent finding [22] that the fluctuation in step spacing on a vicinal Si(111) face are strongly suppressed (with respect to true equilibrium conditions) by the incoming flux. It is thus believed that the origin of the lower number of kinks in growth conditions should be the weaker amplitude of the fluctuation. Obviously, since the weak scaling difference between growth and evaporation is of the same order of magnitude than the error bar, this point should to be checked by complementary and more precise experiments.

**Conclusion**

It is shown that recent improvements of the classical REM technique enable us to study the spiral morphology in perfectly controlled thermodynamics conditions. More precisely the first LODREM images of spirals have been reported. Beyond the qualitative description of the spiral shapes we have shown that the scaling behavior of the distance between consecutive turns of a spiral at growth and at sublimation on Si(111) differs from that predicted from BCF theory but is in good agreement with models taking into account the overlapping diffusion fields of the adatoms on the surface as well as the non instantaneous kinetic attachment at the steps. Moreover we describe how the step movement do not only depend upon super or under saturation conditions but also depend upon the local geometry of the surface.

**Acknowledgements**

The authors are grateful to S. Stoyanov, F. Leroy and D. Kashchiev for helpful discussions.

One of the authors (B.R.) acknowledges the financial support by Sixth Framework Project NANOPHEN (INCO-CT-2005-016696), the hospitality and stimulating working conditions in CRMC-N/CNRS, Luminy, Marseille, France.

## *Figure captions :*

**Fig.1:** *Scheme of the sample holder. Vertical arrows schematise growth conditions when the substrate temperature is lower than the source temperature so that supersaturation conditions are fulfilled ($\tau = T_{source}/T_{substrate} > 1$). The flux leaving the source and impinging the crystal is thus more intense (long arrows) than the desorption flux leaving the substrate (short arrows)*

**Fig. 2.** *Resulting deposited flux $F_{res}$ versus the reduced temperature $1/\tau = T_{substrate}/T_{source}$ plotted for different values of $T_{source}$.*

**Fig. 3:** *Spiral on Si(111) at four different angles of the LODREM screen. Sublimation at 1520 K. Only the common scale parallel to the tilt axis (and thus independent upon the screen angle) is given (screen angles: ~ 30, 40, 80, 85 degrees). Notice that working in grazing incidence enhances the screen defects (white points on the last photography).*

**Fig. 4:** *The central part of the spiral on Si(111) after digital processing (image on the left and the dashed curve on the right) and The Archimedean spiral (solid curve on the right).*

**Fig. 5**: *Non-trivial cases of spirals of sublimation on Si(111) at 1250°C. Vertical black zones correspond to elongated shadow of SiC defects.*

**Fig. 6:** *Direction of the movement of the steps emerging from (a) "negative" spiral hole then (b) "positive" spiral pyramid during growth and sublimation conditions. The dotted line corresponds to the original planar surface.*

**Fig. 7**. *Growth on Si(111) – Dependence of the spiral step distance on the deposition rate*

**Fig 8:** *Sublimation on Si(111) – Dependence of the spiral step distance on the sublimation rate*

# Figures

**Figure 1:**

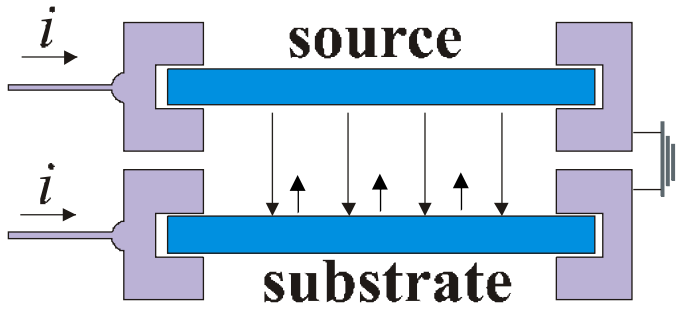

**Figure 2:**

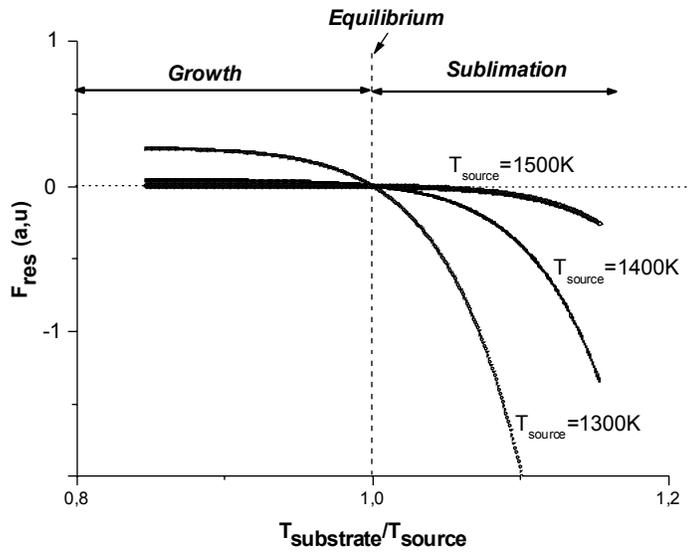

**Figure 3 :**

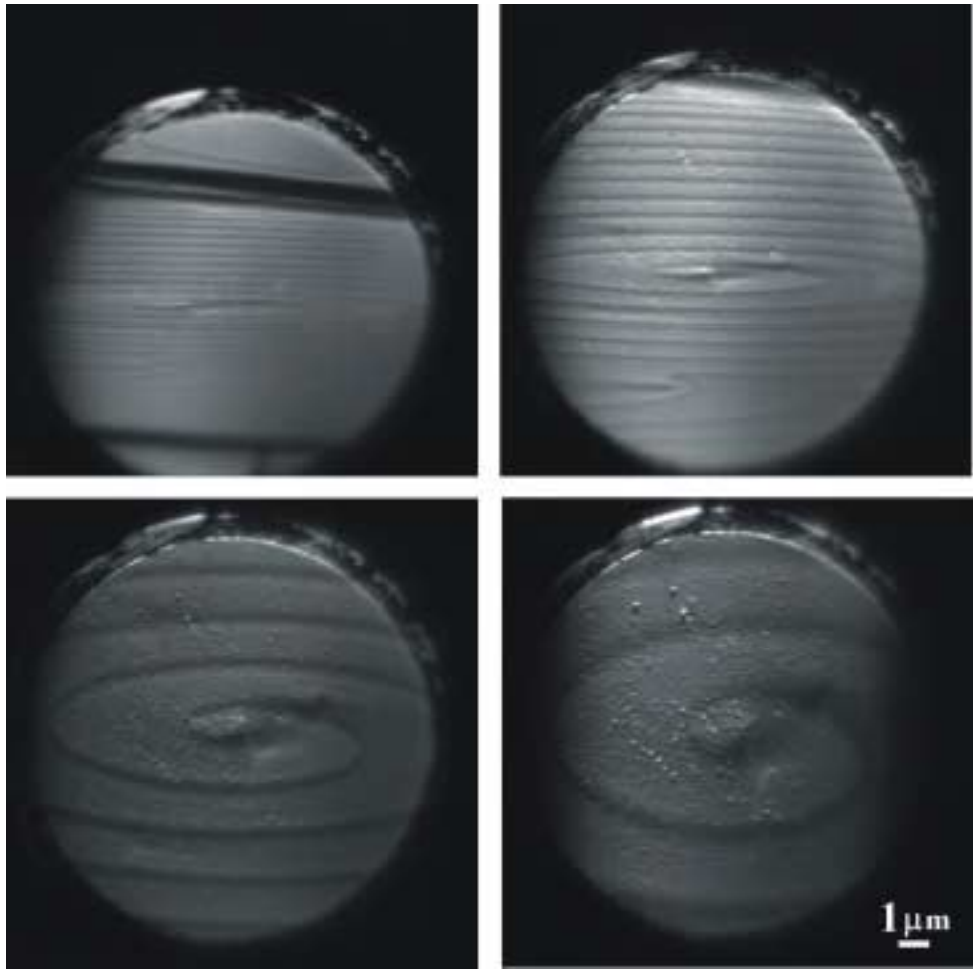

**Figure 4 :**

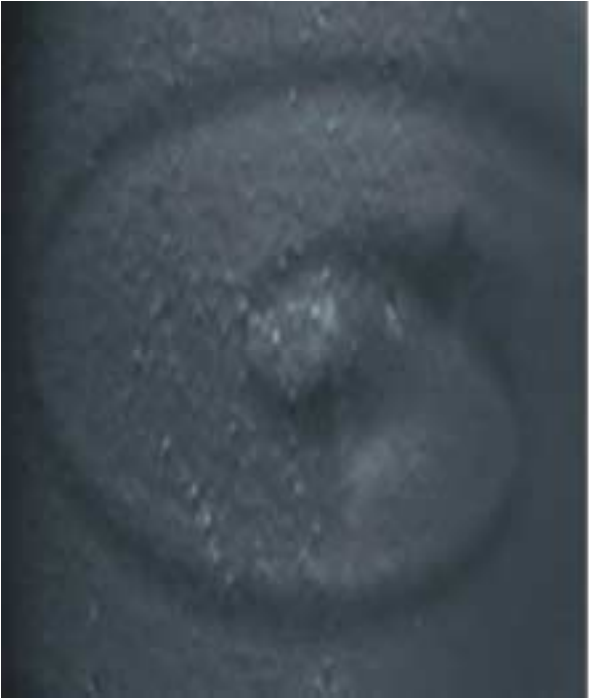 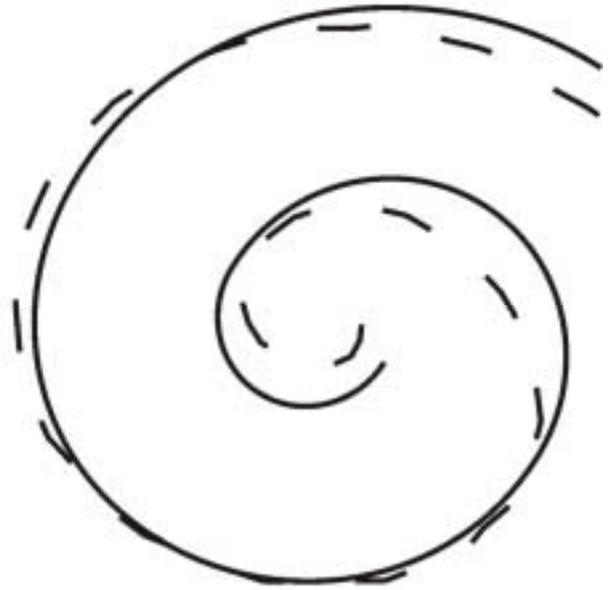

**Figure 5 :**

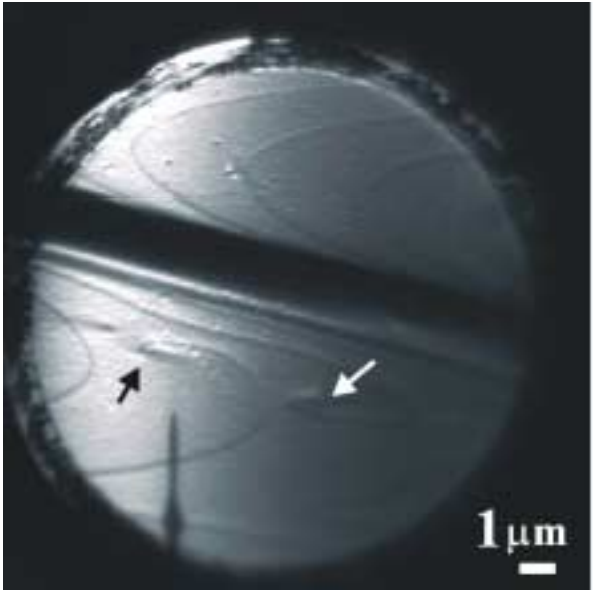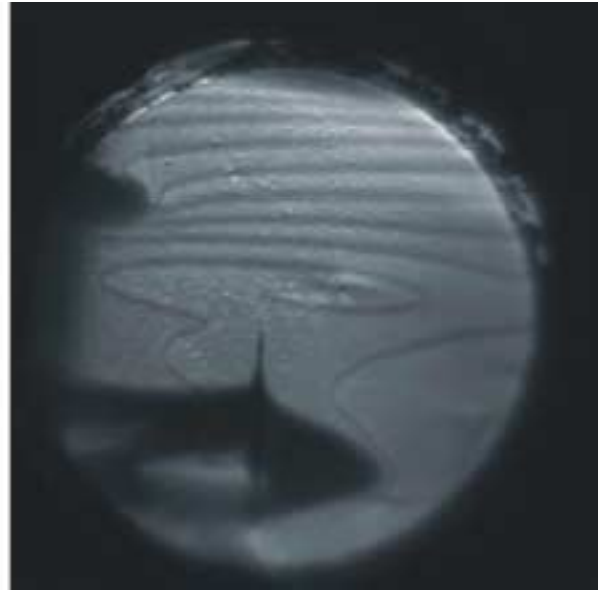

**Figure 6:**

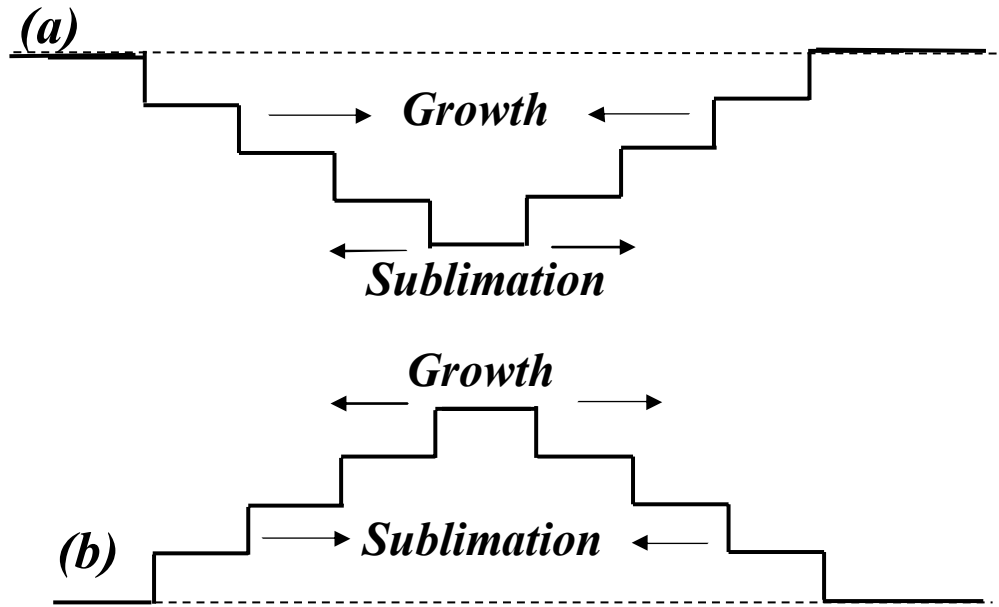

**Figure 7 :**

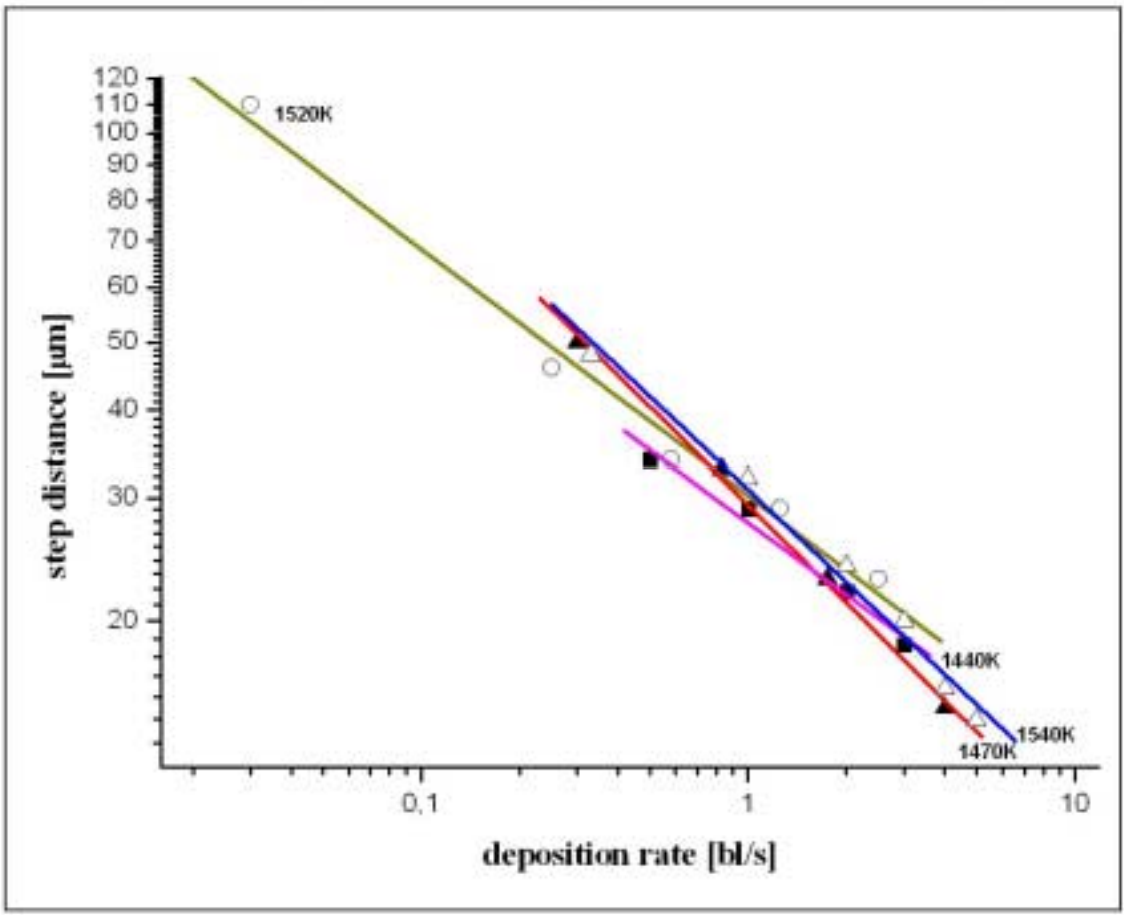

**Figure 8 :**

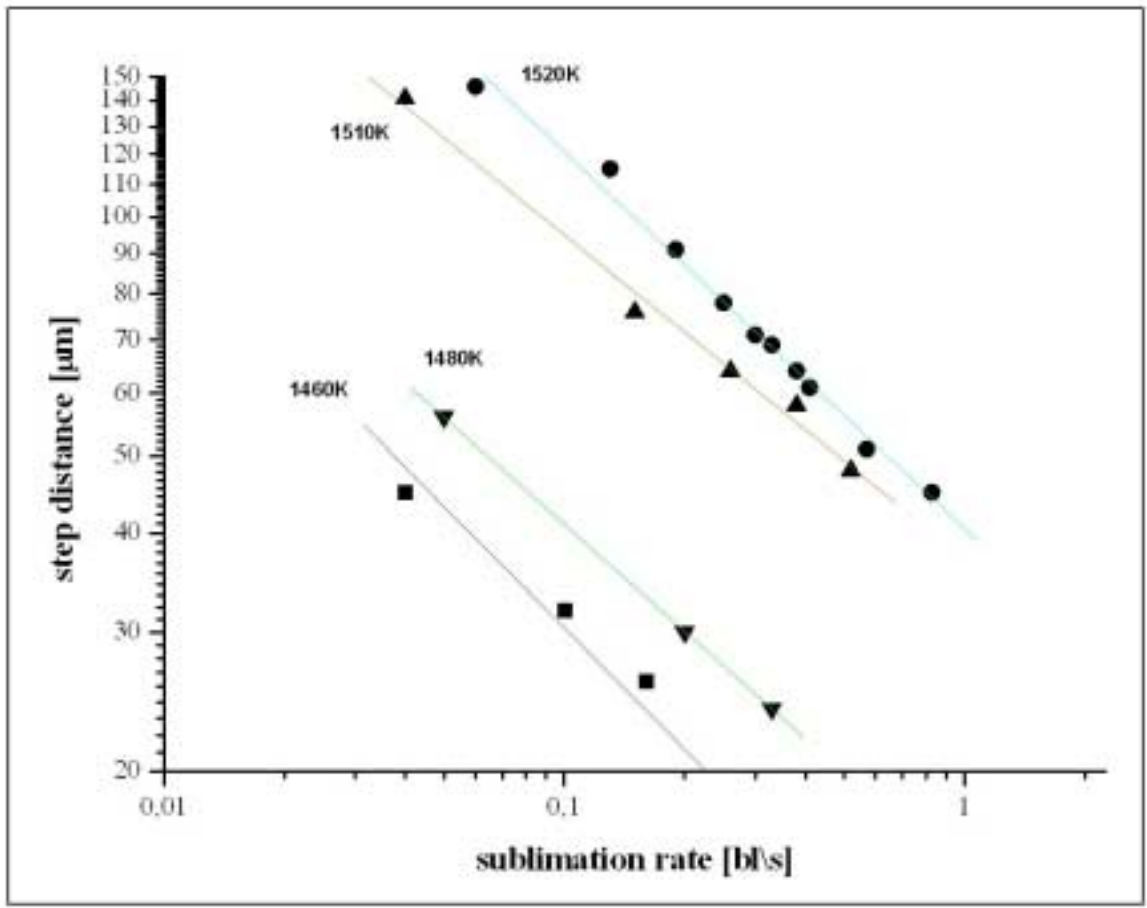